\documentclass[11pt]{article}
\usepackage{amssymb, amsmath, amsthm, epsf, epsfig, color}

\newlength{\originalbase}
\newcommand{\spacing}[1]{\setlength{\baselineskip}{#1\originalbase}}

\begin{document}
\spacing{1.5}
\newtheorem{theorem}{Theorem}[section]
\newtheorem{claim}{Claim}[theorem]
\newtheorem{prop}[theorem]{Proposition}
\newtheorem{remark}[theorem]{Remark}
\newtheorem{lemma}[theorem]{Lemma}
\newtheorem{corollary}[theorem]{Corollary}
\newtheorem{guess}[theorem]{Conjecture}
\newtheorem{conjecture}[theorem]{Conjecture}

\title{Note on Counting Eulerian Circuits}

\author{
Graham R. Brightwell~\thanks{Department of Mathematics, London School
of Economics and Political Science, London, UK}
\and Peter Winkler~\thanks{Bell Labs 2C-365, 700 Mountain Ave.,
Murray Hill, NJ 07974; and Institute for Advanced Study, Princeton
NJ 08540.  Research supported in part by ONR grant N00014-03-M-0141.}
}

\maketitle

\begin{abstract}
We show that the problem of counting the number of Eulerian circuits
in an undirected graph is complete for the class \#P.
\end{abstract}

\medskip

\section{Introduction}

Every basic text in graph theory contains the story of Euler and
the K\"{o}nigsberg bridges, together with the theorem that
guarantees the existence of a circuit traversing every edge of
a graph exactly once, if and only if the graph is connected and
all vertices have even degree.

Later in the text one might find the ``Matrix Tree Theorem'',
which provides an efficient algorithm for counting the number of
spanning trees of a graph; and an application of this and the so-called
``BEST'' Theorem (see below) to count the number of Eulerian circuits
in a {\em directed} graph.  But what about counting Eulerian
circuits in an undirected graph?

This problem is clearly in the class \#P (introduced by Valiant
\cite{V1} in the 1970s), since it is easy to check
whether a candidate circuit traverses each edge once.  Since there
is no known efficient way to count Eulerian circuits, it is natural
to suspect that the problem is \#P-complete, and thus
presumably very difficult---especially in view of Toda's result \cite{T},
which implies that one call to a \#P oracle suffices to solve any
problem in the polynomial hierarchy in deterministic polynomial time.

Researchers have shown a myriad of graphical counting problems to be
\#P-complete, including Hamilton circuits \cite{V2}, acyclic
orientations \cite{L}, and Eulerian orientations \cite{MW1}.
Yet, the complexity of counting Eulerian circuits remained
open (see e.g.\ \cite{J}, Open Problem on p.\ 5, at the end of Section 1.1)
for 25 years---rather mysteriously, especially considering the simplicity
of the reduction below.

Our approach is to show that, with the help of an oracle which counts
Eulerian circuits, a Turing machine can count the number of Eulerian
orientations of any given graph in polynomial time.  The latter
problem was shown to be \#P-hard by Mihail and Winkler \cite{MW1,MW2}
and for completeness we provide a second proof below.

Both reductions proceed by enumeration modulo various primes,
a technique introduced originally by Valiant \cite{V1} and 
utilized later by the authors \cite{BW1,BW2} to settle another
long-open complexity problem, counting the linear extensions of
a partially ordered set.

To show later that we can reconstruct a number uniquely from its values
modulo a small set of primes, it is useful to have a technical lemma 
such as the following.

\begin{lemma}
For any $n\geq 4$, the product of the set of primes
strictly between $n$
and $n^2$ is at least $n!\, 2^n$.
\end{lemma}

\begin{proof}
We use some facts from Hardy and Wright \cite{HW}, Chapter~22, concerning
the functions $\vartheta (n) = \log \prod _{p\leq n} p$, where
$p$ runs over all primes less than $n$, and $\displaystyle \psi(n) =
\sum _{i=1}^{\log n/\log 2} \vartheta (n^{1/i})$. From \cite{HW} we find that
$\vartheta (n) < 2n\log 2$ for $n\geq 1$, and that
$\psi(n)\geq {\frac14}n\log 2$ for $n\geq 2$.

We are interested in the quantity $V=\vartheta (n^2) -\vartheta (n)$.
From the above facts, we have:
\begin{eqnarray*}
V & \geq \psi(n^2) - \sum _{i=2}^{2\log n/\log 2}
\vartheta (n^{2/i}) -\vartheta (n) \\
& \geq {\frac14}n^2\log 2 - \frac{2\log n}{\log 2}\cdot 2n\log 2
- 2n\log 2 \\
& \geq n\log n \geq \log (n!\, 2^n)~,
\end{eqnarray*}
at least provided $n\geq 150$.
The inequality for $4\leq n < 150$ is easily verified by direct calculation.
\end{proof}

\bigskip

It is evident that this lemma is not tight: it is possible to replace the
$n^2$ upper limit by $Kn\log n$, for some suitably large $K$.

\bigskip

\section{Circuits, Orientations, Arborescences and Orbs}

For us a graph $G = \langle V,E \rangle$ will be finite and undirected,
with no loops or multiple edges; if multiple edges are permitted we
use the term {\em multigraph}.  A {\em circuit} $C$ of $G$ is a closed
path, with a direction but no distinguished starting point; it is
{\em Eulerian} if it traverses each $e \in E$ exactly once.  Of course
the possession of even one Eulerian circuit implies all degrees are even,
and it will be convenient for us to denote the degree of a vertex $v$
by $2d_v$ instead of $d_v$.

An {\em Eulerian orientation} of $G$ is an orientation of its edges
with the property that each vertex has the same number (namely,
$d_v$) of incoming and outgoing arcs.  Any Eulerian circuit induces an
Eulerian orientation by orienting each edge in accordance with its
direction of traversal.

If a particular starting edge is chosen for the Eulerian circuit $C$,
originating say at vertex $r$, then $C$ also induces a spanning tree
$T = \{{\rm exit}(v): v \not= r\}$ where exit$(v)$ is the last edge
incident to $v$ used by $C$ before its final return to $r$.  When
oriented according to $C$, $T$ becomes an in-bound spanning tree,
or {\em arborescence}, rooted at $r$.

Let us fix a root $r \in V$ and denote by the term {\em orb} a
pair consisting of an Eulerian orientation and an arborescence
(for that orientation) rooted at $r$.  If an orb is specified,
it is easy to construct a corresponding Eulerian circuit $C$:
simply begin walking from the root $r$, following any unused outgoing arc
from each vertex $v$, except that the tree arc exiting $v$ is avoided
as long as possible.  Since there are $(d_v-1)!$ ways to order the non-tree
outgoing arcs from $v$, and $d_r!$ from $r$, the number of ways to
construct $C$ is precisely $d_r! \prod_{v \not= r} (d_v-1)!$.  However,
this over-counts Eulerian circuits (as we have defined them) by a factor
of $d_r$ since each circuit passes $d_r$ times through $r$.  Hence
orbs and Eulerian circuits are in perfect $\prod_{v \in V} (d_v-1)!$-to-1
correspondence, and thus counting Eulerian circuits is equivalent to
counting orbs.  This result is sometimes known as the ``BEST'' Theorem
after de {\bf B}ruijn, van Aardenne-{\bf E}hrenfest, {\bf S}mith and
{\bf T}utte, although the former two should perhaps get additional credit
as the original discoverers.

Note that for any {\em particular} orientation, one can use the BEST Theorem
together with the Matrix-Tree Theorem \cite{Tu} (due to Tutte, but implicit
in Kirchhoff's work of 1847) to count Eulerian circuits.  The
difficulty in using this approach to count (or even approximate)
the number of Eulerian circuits in $G$ is that the number of
arborescences in an orientation can vary enormously.

\begin{theorem}
Counting Eulerian circuits is \#P-complete.
\end{theorem}

\begin{proof}
It suffices to reduce the problem of counting Eulerian orientations
to counting orbs in a multigraph; the latter is equivalent to counting
orbs in a simple graph since multiple edges can be subdivided without
affecting the number of orbs.  Let us therefore assume that an Eulerian
graph $G = \langle V=\{1,\dots,n\},E \rangle$ has been given, and that
we have an oracle for counting orbs in any multigraph.  We wish to
compute the number $N$ of Eulerian orientations of $G$.

We construct for any odd prime $p$, a graph $G_p$ whose number
of orbs is equivalent to $N$ modulo $p$.  The construction
is shockingly simple: each edge $e$ of $E$ is replaced by $p$
parallel edges $e_1, \dots, e_p$, and a new node $0$ is added
which is adjacent to every node $v$ of $V$ by two parallel edges,
$e(v)$ and $e'(v)$.  We take $0$ as the root of all orbs.

Figure~\ref{fig:g3} shows a small $G$ and the resulting
$G_p$ when $p=3$.

\bigskip

\begin{figure}[here]
\epsfxsize250pt
$$\epsfbox{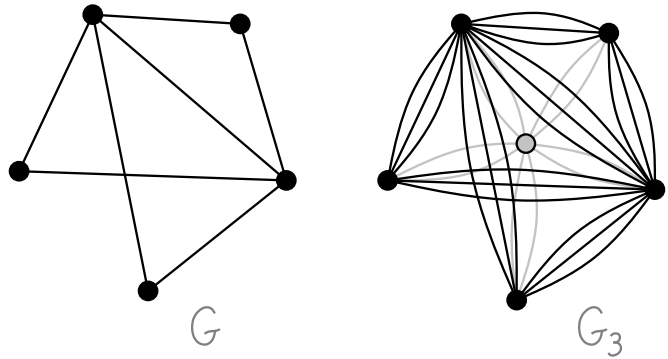}$$
\caption{Reducing {\sc eulerian orientation count} to
{\sc orb count}}\label{fig:g3}
\end{figure}

In $G_p$, the {\em type} $\tau$ of an orb is a function from $E$ to
$\{0,1,\dots,p\} \times \{T,F\}$ which tells how many of 
$e_1, \dots, e_p$ are oriented from the smaller to the larger-numbered
vertex, and whether any is a tree edge (``T'' means ``yes'',
``F'' means ``no'').  A type $\tau$ is {\em special} if 
$\tau(e) \in \{0,p\} \times \{F\}$ for every $e \in E$.

If an orb belongs to a special type, then the common direction of the 
parallel edges corresponding to each $e\in E$ provides an orientation of 
$G$.  This orientation is Eulerian, as otherwise the in-degree and 
out-degree of some vertex of $G_p$ will differ by at least $2p-2$. 
Therefore, in an orb of special type, for each vertex $v$ of $G_p$, 
exactly one of $e(v)$ and $e'(v)$ is directed towards the root~0, and 
this is the arc that carries the edge of the associated tree directed away 
from $v$.  Thus, the number of orbs of special type is precisely 
$2^n \times N$.

On the other hand, we claim that the number of orbs of any
non-special type $\tau$ is a multiple of $p$.  To see this let
$e$ be such that $\tau(e) = (k,X) \not\in \{0,p\} \times \{F\}$.
Suppose first that $0 < k < p$ and $X=F$; then the orbs of type
$\tau$ can be partitioned into $\binom{p}{k}$ equal parts
according to which of $e_1, \dots, e_p$ are oriented from the
smaller to the larger-numbered vertex, and of course
$\binom{p}{k}$ is a multiple of $p$.

If $0 < k < p$ and $X=T$, then the part sizes are multiples
either of $k{p \choose k}$ or $(p\!-\!k){p \choose k}$ since
we must also decide which of the correctly-oriented arcs belongs
to the tree.

Finally, if $k \in \{0,p\}$, and $X=T$, we partition according
to which of the now-parallel arcs is in the tree and there are
$p$ choices.

It follows that the total number of orbs is equivalent to
$2^n \times N$ modulo $p$, and thus we can compute $N \mod p$.
We repeat this process for every prime $p$ between $m$ and
$2m$, where $|E|=m$, and apply the Chinese Remainder Theorem
to nail $N$.
\end{proof}

A similar argument, equally simple, can be used to reduce
{\sc not-all-equal 3-sat count} (shown in \cite{CH} to be
\#P-complete) to {\sc counting eulerian orientations}.
Given an instance of {\sc not-all-equal 3-sat count}, the graph is 
provided with a vertex for each literal and another for each clause, plus 
one spare vertex $s$.  Each clause is connected by a single
edge to its three literals and $s$.  Each literal vertex
is given $p$ parallel edges to its mate (where $p$ is
a prime larger than the number of appearances of any literal),
and a total of $p$ other edges, consisting of some number  
(as mentioned above) to the clauses in which it appears, with the 
remainder going to $s$.

\begin{figure}[here]
\epsfxsize240pt
$$\epsfbox{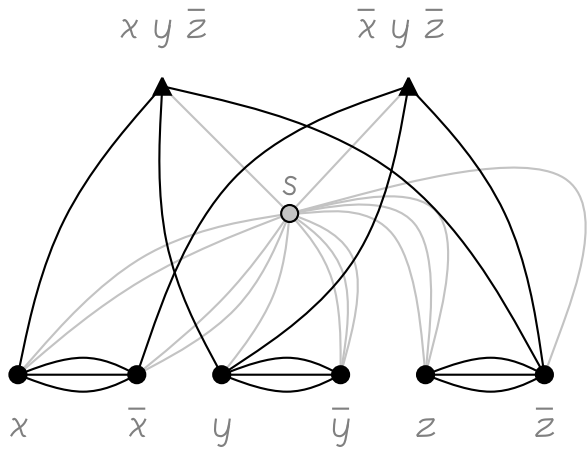}$$
\caption{Reducing {\sc not-all-equal 3sat count} to {\sc eulerian
orientation count}}\label{fig:3sat}
\end{figure}

Figure~\ref{fig:3sat} shows the construction for a particular
2-clause, 3-variable instance with $p=3$.

\bigskip

As in the proof above, orientations which fail to align in parallel
all $p$ edges associated with any given variable fall into classes of
size 0 mod $p$.  Each of the other ``special'' orientations has the 
property that for each variable $x$, either every edge from $x$ to a 
clause vertex points outward and every edge from $\bar x$ to a clause 
vertex points inward, or vice-versa.  Such an orientation corresponds 
to a satisfying assignment, since we cannot have three of the four edges
incident to a clause vertex pointing the same way.  Conversely, given 
a satisfying assignment, we orient all the edges between $x$ and $\bar x$
towards the true literal, orient all other edges incident with a literal 
vertex away from true literals and towards false literals, and orient the 
edge between each clause vertex and $s$ in the necessary manner.  This 
ensures that the in-degree is equal to the out-degree at every vertex 
other than $s$, and therefore also at $s$.  Hence there is an exact 
correspondence between special orientations and satisfying assignments, 
and we proceed as before.

\noindent
{\bf Remark}

Still open is the question of whether there is a fully polynomial
randomized approximation scheme (``fpras'') for counting Eulerian
orientations (as there is, e.g., for Eulerian orientations
\cite{MW2}).  We believe that there is, and even that a particular
Markov chain whose states are orbs and near-orbs mixes rapidly.
We hope and expect that this question will not remain open for
another 25 years.

\end{document}